\documentclass[reprint,amssymb,amsmath,aps,pra,superscriptaddress]{revtex4-1}
\usepackage{graphics}
\usepackage{bbold}
\usepackage{graphicx}
\usepackage{bm}
\usepackage{bbm}
\usepackage{amsmath,amsthm}
\usepackage{lipsum}
\usepackage{amssymb}
\begin{document}
\title{History states of systems and operators}
\author{A.\ Boette}
\affiliation{Departamento de F\'isica-IFLP/CONICET,
	Universidad Nacional de La Plata, C.C. 67, La Plata (1900), Argentina}
\author{R.\ Rossignoli}
\affiliation{Departamento de F\'isica-IFLP/CONICET,
	Universidad Nacional de La Plata, C.C. 67, La Plata (1900), Argentina}
\affiliation{Comisi\'on de Investigaciones Cient\'{\i}ficas (CIC), La Plata (1900), Argentina}

\begin{abstract}
We discuss some fundamental properties of discrete system-time history states. Such states arise for a quantum 
reference clock of finite dimension and lead to a unitary evolution of system states when satisfying a static 
discrete Wheeler-DeWitt-type equation. We consider the general case where system-clock pairs can interact, 
analyzing first their different representations and showing there is always a special clock basis 
for which the evolution for a given initial state can be described by a constant Hamiltonian $H$. 
It is also shown, however, that when the evolution operators form a complete orthogonal set, the history state is  
maximally entangled for any  initial state, as opposed to the case of a constant $H$, and can be generated through a 
simple double-clock setting. We then examine the quadratic system-time entanglement entropy, providing an analytic 
evaluation and showing it satisfies strict upper and lower bounds determined by the energy spread and the  geodesic 
evolution connecting the initial and final states. We finally show that the unitary operator that generates the 
history state can itself be considered as an operator history state, whose quadratic entanglement entropy 
determines its entangling power. Simple measurements on the clock  enable to efficiently determine overlaps 
between system states and also evolution operators at any two times. 
\end{abstract} 

\maketitle

\section{introduction}
The incorporation of time in a fully quantum framework \cite{PaW.83} has recently attracted wide attention 
\cite{Ga.09,GL.15,Mo.14,Ma.15,FC.13,BR.16,Er.17,Pa.17,Ni.18,Ll.18}. On the one hand, it is relevant as a fundamental 
problem and a key issue 
 in the search
for a coherent theory of quantum gravity \cite{DW.67,Ro.04, Ku.11, Is.93,JT.11,Bo.11, Ho.12}. On the other hand, 
 a quantum description of time  enables to exploit the quantum features of superposition and entanglement 
in the development of new models of parallel-in-time simulation \cite{FC.13,BR.16}. 

The concept of time is related to the quantification of evolution through a reference physical system called clock. 
Historically, the readings of this clock provided an external classical parameter, called time. Nonetheless, 
if we aim to introduce time into a fully quantum framework, 
the clock has to be a quantum system itself. This is even more important in attempts to quantize gravity 
where time has to be described by a dynamical entity \cite{Ro.04, Ku.11, Is.93,JT.11, Bo.11, Ho.12}.

Here we describe the system and the reference clock through a discrete system-time history state which  enforces a 
discrete unitary evolution on the system states.  We consider the general case where the system-clock pairs can interact.  
This scenario provides a more general starting point, 
more adequate for some quantum gravity or cosmological models  where interactions between an internal relational 
clock and evolving degrees of freedom cannot be excluded \cite{Bo.11,Ho.12}. 

We first discuss different representations of the history state, showing that   
for a fixed initial state there is always an adequate selection of clock basis for 
which the resultant evolution corresponds to a constant Hamiltonian,   
with the history state  satisfying a  discrete counterpart of a standard Wheeler-DeWitt type equation \cite{DW.67}.  
The general interacting formalism opens, however, new possibilities. The entanglement of the history state is a measure 
of the number of orthogonal states visited by the system at orthogonal times \cite{BR.16}, and for a constant Hamiltonian 
clearly depends  on the seed system state. This dependence becomes, however, attenuated when the  Hamiltonian is not constant 
in time, and in the case where the  evolution operators form a  complete orthogonal set,  it is in fact always {\it maximum}, 
irrespective of the initial state. The corresponding history state  admits, nonetheless,  a simple generation  through a 
two-clock  scenario, where the clocks are linked to conjugate system variables. 
 
  We then analyze the quadratic entanglement entropy of history states, which, as opposed to the standard entropy, 
  can be explicitly evaluated in the general case,  enabling one to   characterize the system evolution  and also  to 
  connect the entanglement of states and operators.  For a general constant Hamiltonian it can be analytically 
  determined for any number of steps. Moreover, we show that it is upper bounded by the quadratic entropy of the energy spread 
  of the initial state and lower bounded by that of the geodesic evolution connecting the initial and final states according 
  to  the Fubini-Study metric \cite{AA.90}. And its average over all initial system states is directly proportional
to the quadratic operator entanglement entropy \cite{Z.00,N.03,P.07,M.13}
 of the unitary gate that generates the history state. Through the channel-state 
  duality \cite{Ja.72,Ch.75,Gr.05,Du.05,Mi.13}, it is also shown that the pure state which represents the latter  
  is itself an operator history state, whose quadratic entanglement entropy determines its  entangling power. 
   
Finally, we show that through measurements on the clock it is possible to use both system and operator history states 
to efficiently determine the overlap between system states  and also the trace of the evolution operator between 
any two-times. The latter reduces to the trace of a unitary operator (result of the DQC1 circuit \cite{KL.98}) 
for the simple case of a qubit clock. The properties of general discrete history states and their entanglement are 
discussed in section \ref{II}, whereas the entanglement and history states 
of unitary operators are discussed in \ref{III}. Conclusions are finally given in \ref{IV}. 

\section{Discrete history states\label{II}}
We consider a system $S$ and a reference clock system $T$ in a joint pure state $|\Psi\rangle\in {\cal H}_S\otimes {\cal H}_T$, 
with ${\cal H}_T$ of finite dimension $N$. Any such state can be written as 
\begin{eqnarray} 
|\Psi\rangle&=&{\textstyle\frac{1}{\sqrt{N}}}
\sum_{t} |S_t\rangle|t\rangle\,,\label{St1}\end{eqnarray}
where $|t\rangle$, $t=0,\ldots,N-1$, are orthogonal states of  $T$  ($\langle t|t'\rangle=\delta_{tt'}$) and $|S_t\rangle$ 
are states of $S$, not necessarily orthogonal or normalized, yet satisfying $\sum_{t}\langle S_t|S_t\rangle/N=\langle \Psi|\Psi\rangle=1$. 
Consider now a unitary operator ${\cal U}$ for the whole system  of the form 
\begin{equation}
    {\cal U}=\sum_{t=1}^N U_{t,t-1}\otimes |t\rangle\langle t-1|\,,
\label{Upsi}
\end{equation}
where $t=N$ is identified with $t=0$ and $U_{t,t-1}$ are arbitrary unitary operators on $S$ 
satisfying $U_{0,N-1}\ldots U_{1,0}=\mathbb{1}$. If  $|\Psi\rangle$ fulfills the eigenvalue equation
\begin{equation}
    {\cal U}|\Psi\rangle=|\Psi\rangle\,,\label{Ueig}
\end{equation}
 the states $|S_t\rangle$ will undergo a {\it unitary} evolution with $t$: 
\begin{eqnarray}
  |S_t\rangle=\sqrt{N}\langle t|\Psi\rangle&=&\sqrt{N}\langle t|{\cal U}|\Psi\rangle\nonumber\\&=&
  U_{t,t-1}|S_{t-1}\rangle\label{Ux1}=U_t|S_{0}\rangle\,,\label{utw}
\end{eqnarray}
where $U_t=U_{t,t-1}\ldots U_{1,0}$, 
with $U_0=\mathbb{1}$. The states $|S_t\rangle$  will then have a unit norm if $|\Psi\rangle$ is normalized.  

Thus, the  state (\ref{St1}) is a  discrete finite dimensional version  of the  history state of 
the Page-Wootters formalism \cite{PaW.83,GL.15}. Moreover,   
writing ${\cal U}=\exp[-i{\cal J}]$, with ${\cal J}$  hermitian 
(and spectrum $\subset[0,2\pi)$),  Eq.\ (\ref{Ueig}) is equivalent to   
\begin{equation}
   {\cal J}|\Psi\rangle=0\label{WDW}\,, 
\end{equation} 
which is a  discrete cyclic version of a Wheeler-DeWitt type equation \cite{DW.67}. Note, however, 
that ${\cal J}$ will contain  $S-T$ interaction terms in the general case where  $U_{t,t-1}$ depends on $t$.  

A unitary evolution of the states $|S_t\rangle$ actually occurs if $|\Psi\rangle$ is  {\it any} eigenstate of ${\cal U}$: 
Its eigenvalues are $e^{-i2\pi k/N}$, $k=0,\ldots,N-1$, and its eigenstates have all the form (\ref{St1}) with $|S_t\rangle$ 
satisfying a {\it shifted} unitary evolution: $|S_t\rangle=e^{i2\pi k/N}U_{t,t-1}|S_{t-1}\rangle=e^{i2\pi kt/N}U_t|S_0\rangle$. 
Each eigenvalue has degeneracy equal to the dimension $d_S={\rm dim}{\cal H}_S$ of the system space, with its eigenspace spanned by  
 orthogonal history states $|\Psi^l_k\rangle$ generated by $d_S$ orthogonal initial states
  $|S_{0}^l\rangle$: $\langle \Psi^l_k|\Psi^{l'}_{k'}\rangle=\langle S_0^l|S_0^{l'}\rangle=\delta^{ll'}$  \cite{BR.16}. 

If $U_{t,t-1}$ is independent of $t$ 
$\forall$ $t=1,\ldots,N$, then    
\begin{equation}U_{t,t-1}=\exp[-iH_S]\,,\label{UHS}
\end{equation}
with $H_S$ a fixed hermitian Hamiltonian for system $S$ with eigenvalues $2\pi k/N$,  $k$ {\it integer}. 
The operator (\ref{Upsi}) becomes then {\it separable}: ${\cal U}=\exp[-i H_S]\otimes \exp[-iP_T]$, implying 
\begin{equation}
{\cal J}=H_S\otimes \mathbbm{1}+\mathbbm{1}\otimes P_T\,,\label{J}\end{equation}
which contains no interaction terms. Here $P_T$ is the generator of time translations, 
satisfying $e^{-iP_T}|t-1\rangle=|t\rangle$ $\forall$ $t$ and 
$P_T|k\rangle_T=\frac{2\pi k}{N} |k\rangle_T$,  with  $|k\rangle_T$ the discrete Fourier transform (DFT) of the states $|t\rangle$: 
\begin{equation}|k\rangle_T=\frac{1}{\sqrt{N}}\!\sum_t e^{i2\pi kt/N}|t\rangle\,,\;\;k=0,\ldots,N-1\,.\label{DFT}\end{equation} 
Eqs.\ (\ref{WDW})--(\ref{J}) then become an exact discrete version of the usual static Wheeler-DeWitt equation \cite{GL.15}. 
The ensuing condition $\langle t|{\cal  J}|\Psi\rangle=0$ implies 
\begin{equation}-\langle t|P_T|\Psi\rangle=H_S|S_t\rangle\;
\,,\end{equation}
which  is a  discrete version of Schr\"odinger's equation: As  $-\langle t|P_T|t'\rangle=i\frac{\partial}
{\partial t}\frac{1}{N}\sum_{k}e^{i2\pi k(t-t')/N}$,  
for $N\rightarrow\infty$  $-\langle t|P_T|t'\rangle\rightarrow i\delta'(t-t')$ 
and  $-\langle t|P_T|\Psi\rangle\rightarrow i\frac{\partial}{\partial t}|S_t\rangle$. 

 \subsection{Representations and entanglement of the history state}
By considering an arbitrary orthogonal basis $\{|q\rangle\}$ 
of ${\cal H}_S$, we may first rewrite  $|\Psi\rangle$ as  
\begin{equation}
|\Psi\rangle=\frac{1}{\sqrt{N}}\sum_{q,t}\psi(q,t)
|qt\rangle\label{St2}\,,    \end{equation}
where $|qt\rangle=|q\rangle|t\rangle$ and  $\psi(q,t)=\langle q|S_t\rangle=\sqrt{N}\langle qt|\Psi\rangle$ is a 
``wave function'' satisfying  a unitary evolution with $t$: $\psi(q,t)=\sum_{q'}\langle q|U_{t,t-1}|q'\rangle\psi(q',t-1)$. 

We may then obtain the Schmidt decomposition of $|\Psi\rangle$, which we will  here write as 
 \begin{equation}|\Psi\rangle=\sum_k \lambda_k\,|k\rangle_S\,|-k\rangle_T\label{Scm}\,,\end{equation}
where $\lambda_k>0$ are the singular values of the matrix $\psi(q,t)/\sqrt{N}$  and $|k\rangle_{S(T)}$ orthonormal 
states of $S$ ($T$) derived from the singular value decomposition of $\psi(q,t)$, with $|-k\rangle\equiv|N-k\rangle$.
 They are eigenstates of the reduced states $\rho_{S(T)}={\rm Tr}_{T(S)}\,|\Psi\rangle\langle\Psi|$, with 
$\lambda_k^2$ their non-zero eigenvalues. While the states $|S_t\rangle\propto \langle t|\Psi\rangle$ are not necessarily 
orthogonal but are equally probable, the states $|k\rangle_S\propto\, {_T}\langle-k|\Psi\rangle$ are all orthogonal but 
not equally probable, with  $\lambda_k^2$ representing a ``permanence'' probability.

In the constant case (\ref{UHS})--(\ref{J}), the Schmidt states  $|k\rangle_S$ and $|k\rangle_T$ are 
just the eigenstates of $H_S$ and $P_T$:
\begin{equation} 
H_S|k\rangle_S=\frac{2\pi k}{N}|k\rangle_S\,,\;\;P_T|k\rangle_T=\frac{2\pi k}{N}|k\rangle_T\,,
\label{KST}\end{equation}
since $|S_t\rangle=e^{-i H_S t}|S_0\rangle=\sum_{k} \lambda_k e^{-i2\pi k t/N}|k\rangle_S$ with 
$\lambda_k={_{S}}\langle k|S_0\rangle$, and hence  
$|\Psi\rangle=\frac{1}{\sqrt{N}}\sum_{k,t}\lambda_k e^{-i2\pi kt/N}|k\rangle_S |t\rangle$ becomes Eq.\ (\ref{Scm}), 
 with $|k\rangle_T$ the strictly orthogonal states (\ref{DFT}).  
 The Schmidt coefficients $\lambda_k$ represent in this case  the  distribution of $|S_0\rangle$ over distinct energy eigenstates 
 (in case of degeneracy, $\lambda_k|k\rangle_S$ denotes the projection of $|S_0\rangle$ onto  
 the eigenspace of energy $2\pi k/N$ ($mod\, 2\pi$),  with  $\lambda_k^2$ the total probability of measuring this  
 energy in $|S_0\rangle$).  It is then apparent from Eqs.\  (\ref{J}) and (\ref{Scm}) that $|\Psi\rangle$ 
 satisfies Eq.\ (\ref{WDW}), which becomes a  zero  ``total momentum'' condition: $k_S+k_T=0$ ($mod\, N$).

\begin{figure}[h!]
 \centering
 \includegraphics{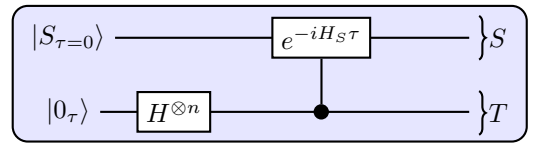}
 \caption{Schematic circuit representing the generation of the {history state} (\ref{St22a}) in the special time basis, 
 	where the system evolves according to a  constant  Hamiltonian $H_S$. Here $H^{\otimes n}$ 
 	denotes the Hadamard operator over $n$ qubits, with $2^n=N$.} \label{f1}
\end{figure}

In the case of arbitrary unitary operators $U_{t,t-1}$ in (\ref{Upsi}), for any  given initial state $|S_0\rangle$  there is always, 
however, a special  orthogonal  basis of ${\cal H}_T$   for which  the corresponding states of $S$ {\it evolve according 
	to a  constant Hamiltonian $H_S$} satisfying (\ref{KST}). It is just necessary to use the inverse DFT 
of the Schmidt states $|k\rangle_T$ of (\ref{Scm}), 
\begin{equation}
|\tau\rangle={\textstyle\frac{1}{\sqrt{N}}}\sum_k\,e^{-i2\pi k\tau/N}|k\rangle_T\,,\label{tau}
\end{equation}
with $k,\tau=0,\ldots,N-1$ (if the Schmidt rank 
 is less than $N$, the states $|k{\rangle}_T$ of  (\ref{Scm}) can be completed with orthogonal states),  
 which will not coincide in general with the original states $|t\rangle$. The state (\ref{Scm})  then becomes 
\begin{equation}
|\Psi\rangle={\frac{1}{\sqrt{N}}}
	\sum_{\tau,k}\lambda_k\,e^{-i2\pi k\tau/N}|k\rangle_S|\tau\rangle=
	\frac{1}{\sqrt{N}}\sum_{\tau}|S_{\tau}\rangle|\tau\rangle
	\label{St22a}\,,\end{equation}
where $|S_\tau\rangle=\sum_k e^{-i2\pi k\tau/N}\lambda_k|k\rangle_S$ satisfies 
\begin{eqnarray}|S_\tau\rangle=\sqrt{N}\langle \tau|\Psi\rangle&=&
\exp[-i\tau H_S]|S_{\tau=0}\rangle\,,\label{HStau}\end{eqnarray}
with $|S_{\tau=0}\rangle=\sum_{k}\lambda_k\,|k\rangle_S$ 
and $H_S$ defined over the Schmidt states $|k\rangle_S$ by Eq.\ (\ref{KST}). The Schmidt coefficients $\lambda_k$ can then be  interpreted 
as the distribution of $|S_{\tau=0}\rangle$ over these energy eigenstates. In terms of the operators 
$H_S$ and $P_T$ defined by (\ref{KST}), $|\Psi\rangle$ satisfies  Eq.\ (\ref{WDW}) also for an effective non-interacting 
${\cal J}$ of the form (\ref{J}), and can be generated from $|S_{\tau=0}\rangle|0_\tau\rangle$ with the circuit of Fig.\ (\ref{f1}). 
 
Assuming now $d_S= N$ (the Schmidt decomposition selects in any case subspaces of equal dimension on $S$ and $T$) we can also 
consider the inverse DFT of the system Schmidt states,  $|\xi\rangle=\frac{1}{\sqrt{N}}\sum_k e^{-i2\pi k \xi/N}|k\rangle_S$, 
which satisfy $e^{-iH_S}|\xi\rangle=|\xi+1\rangle$   
and are the special system states analogous to $|\tau\rangle$.  We can  then also rewrite $|\Psi\rangle$ as 
\begin{eqnarray}
|\Psi\rangle&=&{\textstyle\frac{1}{\sqrt{N}}}
\sum_{\xi,\tau}\Lambda_{\xi-\tau}|\xi\tau\rangle
=\sum_{\xi}\Lambda_\xi|\Psi_\xi\rangle
\label{Psit}\,,
\end{eqnarray}
where $\sqrt{N}\langle \xi\tau|\Psi\rangle=\Lambda_{\xi-\tau}$ depends just on $\xi-\tau$,  
and  
\begin{equation}\Lambda_\xi=\frac{1}{\sqrt{N}}\sum_k e^{i2\pi k \xi/N}\lambda_k\,,\end{equation} 
is the DFT  of the 
Schmidt coefficients $\lambda_k$, with  $|\Psi_\xi\rangle=\frac{1}{\sqrt{N}}\sum_\tau|\xi+\tau\rangle|\tau\rangle$ 
orthogonal 
maximally entangled  history states: $\langle\Psi_\xi|\Psi_{\xi'}\rangle=\delta_{\xi\xi'}$ 
($|\xi+\tau\rangle\equiv|\xi+\tau-N\rangle$ if $\xi+\tau\geq N$). 

The representation (\ref{Psit}) is then  ``conjugate'' to  (\ref{Scm}), expressing $|\Psi\rangle$ as a superposition of 
 maximally entangled orthogonal history states. Like (\ref{Scm}), it is {\it symmetric} in $S-T$: States 
 $|S_\tau\rangle=\sqrt{N}\langle \tau|\Psi\rangle=\sum_\xi\Lambda_{\xi-\tau}|\xi\rangle$ evolve unitarily with $\tau$ 
 (Eq.\ (\ref{HStau})) while clock states $|T_\xi\rangle=\sqrt{N}\langle\xi|\Psi\rangle=\sum_{\tau}\Lambda_{\xi-\tau}|\tau\rangle$ evolve 
 unitarily with $\xi$: 
 \begin{eqnarray}|T_\xi\rangle=\sqrt{N}\langle\xi|\Psi\rangle&=&
\exp[-i\xi P_T]|T_{\xi=0}\rangle\,,\label{PTxi}\end{eqnarray}
 where $|T_{\xi=0}\rangle=\sum_k \lambda_k|-k\rangle_T$, 
 complementing Eq.\ (\ref{HStau}). Both $\xi$ and $\tau$ always run from $0$ to $N-1$ with uniform weight, irrespective of the seed state. 
 
From the Schmidt decomposition (\ref{Scm}) we can evaluate the system-time 
entanglement entropy \cite{BR.16}
\begin{equation}
E(S,T)=S(\rho_S)=S(\rho_T)=-\sum_k \lambda_k^2\log_2 \lambda_k^2\label{S}\,,
\end{equation}
where $S(\rho)=-{\rm Tr}\,\rho\log_2\rho$. If $|S_0\rangle$ happens to be a common eigenstate of all $U_{t,t-1}$, 
such that  $|S_t\rangle=e^{-i\phi_t}|S_0\rangle$ 
$\forall$ $t$, then $|\Psi\rangle\propto|S_0\rangle\sum_t e^{-i\phi_t}|t\rangle$ becomes separable and $E(S,T)=0$ (stationary state), 
whereas  if 
all $|S_t\rangle$ are orthogonal (i.e.\ fully distinguishable), $|\Psi\rangle$ becomes maximally entangled, with (\ref{St1}) 
already the Schmidt decomposition and $E(S,T)=\log_2 N$  maximum.  Thus, $2^{E(S,T)}$  measures the actual system evolution time, 
in the sense of counting the number of effective equally probable orthogonal states the system visits at 
orthogonal times. For  constant $U_{t,t-1}$ (Eq.\ (\ref{UHS})),  $E(S,T)$ is just a measure of the  {\it energy  spread} ($mod\,2\pi$) 
of the initial state, as $\lambda_k= {_S}\langle k|S_0\rangle$. A similar interpretation holds for the general case in terms 
of the effective $H_S$ defined by (\ref{KST}).  

On the other hand, the entropy determined by the conjugate distribution $|\Lambda_\xi|^2$, 
\begin{equation}\tilde{E}(S,T)=-\sum_\xi |\Lambda_\xi|^2\log_2|\Lambda_\xi|^2\,, \label{Sc}
\end{equation}
measures the spread of $|\Psi\rangle$ over maximally entangled evolutions, or equivalently, the spread of system states 
$|\xi\rangle$ for a given clock state $|\tau\rangle$ (or viceversa), and is a measure of {\it time uncertainty}. 
It vanishes when $|\Psi\rangle$ is maximally entangled ($\Lambda_\xi=\delta_{\xi,0}$ if  
$\lambda_k=\frac{1}{\sqrt{N}}$ $\forall\,k$), in which case there is complete synchronization 
between the special system and clock basis states ($|\Psi\rangle=\frac{1}{\sqrt{N}}\sum_\tau |\tau\rangle|\tau\rangle)$, 
and becomes maximum for a product  state ($\Lambda_\xi=\frac{1}{\sqrt{N}}$ $\forall$ $\xi$ if 
$\lambda_k=\delta_{k,0}$), in which case system and clock states  are completely uncorrelated, as seen from (\ref{Psit}).  
 These two entropies satisfy the entropic uncertainty relation  
\cite{BR.16} (see also \cite{DCT.91,PDO.01,Hi.57,Ll.18})
\begin{equation} E(S,T)+\tilde{E}(S,T)\geq \log_2\,N\,,
\end{equation}
which is saturated in the previous limits.

\subsection{The case of a complete set of evolution operators\label{IIbb}}
While for a constant Hamiltonian the  system-time entanglement (\ref{S}) clearly 
depends on the seed  state $|S_0\rangle$, such dependence becomes softened in the more 
general case where the operators $U_{t,t-1}$ depend on $t$ and do not commute 
among themselves, i.e.\ when the `Hamiltonian' $H_t\propto \ln U_{t,t-1}$ is 
time-dependent and $[H_t,H_{t'}]\neq 0$ for some pairs $t\neq t'$. If they  
have no common eigenstate,  $|\Psi\rangle$ will  be entangled for {\it any} $|S_0\rangle$. 
The extreme case is that  where the $U_t$'s of (\ref{utw})   
form a {\it complete} set of {\it orthogonal} unitaries on $S$, such that  
\begin{equation}{\rm Tr}\,[U_t^\dagger U_{t'}]=d_S\delta_{tt'}\,,\;\;t,t'=0,\ldots,d_S^2-1,\label{Uto}\end{equation}
implying $N=d_S^2$. In this case the history state (\ref{St1}) 
becomes  {\it maximally entangled} for {\it any} initial state $|S_0\rangle$:  
\begin{equation} E(S,T)=\log_2 d_S\,,\label{Eds}\end{equation}
such that $|\Psi\rangle=\frac{1}{\sqrt{d_S}}\sum_k |k\rangle_S|-k\rangle_T$ $\forall$ $|S_0\rangle$.  \\   
{\it Proof:}   We may view Eq.\ (\ref{Uto}) as the scalar product between column vectors $\frac{1}{\sqrt{d_S}}\bm{U}_t$ 
of a $d_S^2\times d_S^2$ unitary matrix 
${\bm U}$ of elements ${\bm U}_{ij,t}=\frac{1}{\sqrt{d_S}}\langle i|U_t|j\rangle$, with $\{|i\rangle\}$ 
any orthonormal basis of $S$, 
such that (\ref{Uto}) is equivalent to ${\bm U}^\dagger{\bm U}=\mathbbm{1} _{d_S^2}$. This matrix then satisfies as well 
${\bm U}{\bm U}^\dagger=\mathbbm{1}_{d_S^2}$, i.e.\ 
$\sum_{t}\langle i|U_t|j\rangle\langle l|U_t^\dagger|k\rangle=d_S\delta_{ik}\delta_{jl}$, which implies
$\sum_t U_t|j\rangle \langle l|U_t^\dagger=d_S\delta_{jl}\mathbbm{1}_{S}$ and hence 
\begin{equation} 
\sum_t U_t|S_0\rangle\langle S_0'| U_t^\dagger=d_S\,\langle S_0'|S_0\rangle\,\mathbbm{1}_{S}\label{res}\,,
\end{equation} 
for any two states $|S_0\rangle$, $|S_0'\rangle$ of $S$. In particular, 
for $|S_0\rangle=|S_0'\rangle$, Eq.\ (\ref{res}) implies  a {\it maximally mixed}  
reduced state $\rho_S={\rm Tr}_T|\Psi\rangle\langle\Psi|$ for {\it any} seed state $|S_0\rangle$: 
\begin{equation}
\rho_S=\frac{1}{d_S^2}\sum_t U_t|S_0\rangle\langle S_0|U_t^\dagger=\frac{1}{d_S}\mathbbm{1}_S\,.\label{rhosm}
\end{equation}
 Eq.\ (\ref{rhosm}) then leads to Eq.\ (\ref{Eds}). \qed 

Therefore, a complete orthogonal set of $U_t$'s ensures that the system will visit $d_S$ orthogonal states 
irrespective of the initial state $|S_0\rangle$. The Schmidt decomposition (\ref{Scm}) 
will then select a subspace of ${\cal H}_T$ of
 dimension $d_S$ connected with $S$ through $|\Psi\rangle$. Due to the $d_S$-fold degeneracy 
$\lambda_k=\frac{1}{\sqrt{d_S}}$ $\forall$ $k$, any orthogonal basis $\{|k\rangle_T\}$ 
of this subspace can be used in (\ref{Scm}), with 
all states $|k\rangle_S=\sqrt{d_S}\,{_T}\langle-k|\Psi\rangle$ directly orthogonal. 

A convenient choice of complete orthogonal set is provided by the Weyl operators \cite{W,Ga.88,Er.16}
\begin{equation}
U_{t}\equiv U_{pq}= \exp[i 2\pi pQ/d_S]\exp[-i 2\pi q P/d_S]\,,\label{UW}
\end{equation}
where $p,q=0,\ldots,d_S-1$,  $t=qd_S+p$, $Q|q\rangle=q|q\rangle$, $P|p\rangle=p|p\rangle$ 
and $\{|q\rangle\}$, $\{|p\rangle\}$ are orthogonal 
bases of $S$ related through a DFT: $|p\rangle=\frac{1}{\sqrt{d_S}}\sum_q e^{i2\pi pq/d_S}|q\rangle$.  
They satisfy, for any eigenstate $|q_0\rangle$ of $Q$, 
\begin{equation}U_{pq}|q_0\rangle=e^{i2\pi p(q_0+q)/d_S}|q_0+q\rangle\end{equation}
which implies Eq.\ (\ref{Uto}), i.e.\ ${\rm Tr}\,U_{p'q'}^\dagger U_{pq}=d_S\delta_{q'q}\delta_{p'p}$. 

The discrete evolution under these operators can then be achieved by application of 
just {\it two} different unitaries $U_{t,t-1}$  to the preceding 
state (here $m\geq 1$, integer):
\begin{equation} U_{t,t-1}=\left\{\begin{array}{lr}e^{i2\pi Q/d_S}&t\neq m d_S\\
e^{-i2\pi P/d_S}e^{i2\pi Q/d_S}&t=m d_S\end{array}\right.\,.\label{MS}
\end{equation}
 For instance, if $S$ is a qubit ($d_S=2$) we 
 may take 
$Q=(\mathbbm{1}-\sigma_z)/2$, $P=(\mathbbm{1}-\sigma_x)/2$, with $e^{i2\pi Q/d_S}=\sigma_z$, $e^{-i2\pi P/d_S}=\sigma_x$.  
Hence,  $|\Psi\rangle=
\frac{1}{2}[|S_0\rangle|0\rangle+\sigma_z|S_0\rangle|1\rangle+
\sigma_x|S_0\rangle|2\rangle+i\sigma_y|S_0\rangle|3\rangle]$ is maximally entangled  $\forall$ $|S_0\rangle$ ($E(S,T)=1$),  
with $|S_1\rangle=\sigma_z|S_0\rangle$, $|S_2\rangle=\sigma_x|S_0\rangle=
-i\sigma_y|S_1\rangle$,  $|S_3\rangle=i\sigma_y|S_0\rangle=\sigma_z|S_2\rangle$ 
and $|S_0\rangle=-i\sigma_y|S_3\rangle$. 

\begin{figure}[h!]
 \includegraphics[scale=0.8]{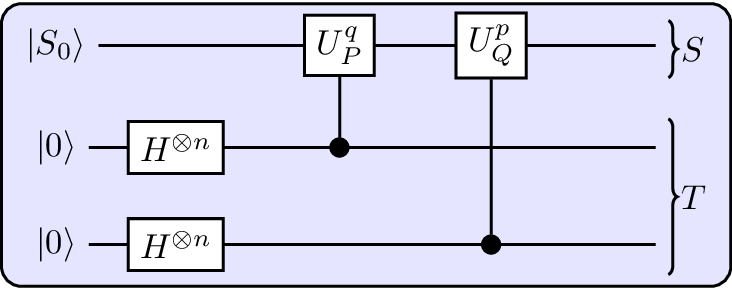}
 \caption{Schematic circuit representing the generation of a {maximally entangled} history state $|\Psi\rangle$, 
 	for any initial system state $|S_0\rangle$.  Here $U_P=e^{-i2\pi P/d_S}$, $U_Q=e^{i2\pi Q/d_S}$, with $P,Q$ 
 	conjugate operators  on $S$ and $2^n=d_S$.} \label{f2}
\end{figure}

In the general case, it is here natural to view system $T$ as formed by two clocks with identical 
Hilbert space dimension $d_S$, which govern 
{\it time-independent} Hamiltonians $H_1=-2\pi  Q/d_S$ and $H_2=2\pi P/d_S$ associated with conjugate 
operators $Q$, $P$ on $S$. Then  
we may write the history state (\ref{St1}) for the operators (\ref{UW}) as 
\begin{equation}
|\Psi\rangle=\frac{1}{d_S^2}\sum_{p,q}U_{pq}|S_0\rangle|p\rangle_{T_1}|q\rangle_{T_2}\,,
\end{equation}
which represents a history state of history states. It can then be implemented with the circuit of Fig.\ \ref{f2}. 

\subsection{The quadratic $S-T$ entanglement entropy: Analytic evaluation and bounds\label{IIc}}
The analytic evaluation of the entropy (\ref{S}) in the general case requires the 
determination of the singular values  $\lambda_k$, i.e., 
the eigenvalues $\lambda_k^2$ 
of $\rho_S$ or $\rho_T$, which is difficult in most cases. It is then convenient 
to use the quadratic (also called linear) 
entropy $S_2(\rho)=2{\rm Tr}[\rho(\mathbbm{1}-\rho)]=2(1-{\rm Tr}\,\rho^2)$,  
which does not require explicit knowledge of the eigenvalues and is a linear  function of the purity ${\rm Tr}\,\rho^2$. 
Like $S(\rho)$, it vanishes iff $\rho$ is pure and is maximum iff $\rho$ is 
maximally mixed (with $S_2(\rho)=1$ for a maximally 
mixed single qubit state), satisfying the majorization relation 
$S_2(\rho')\geq S_2(\rho)$ if $\rho'\prec \rho$ \cite{Bha.97,CR.03}. 
The associated $S-T$ entanglement entropy is 
\begin{eqnarray}
E_2(S,T)&=&S_2(\rho_S)=S_2(\rho_T)=2(1-\sum_k \lambda_k^4)\label{S21}\\
&=&2(1-{\textstyle\frac{1}{N^2}}\sum_{t,t'}|\langle S_t|S_{t'}\rangle|^2)\,,\label{S22}
\end{eqnarray}
and can be determined just from the overlaps between the evolved states. 
For the complete orthogonal set (\ref{Uto}), it is easily  verified that $\sum_{t,t'}|\langle S_t|S_{t'}\rangle|^2=d_S^3$, so that 
$E_2(S,T)=2(1-\frac{1}{d_S})$ becomes maximum. 

The overlaps $\langle S_t|S_{t'}\rangle$ are also experimentally accessible 
through a measurement at the clock $T$ of the non-diagonal operators 
$|t'\rangle\langle t|$ ($t\neq t'$): 
\begin{equation}\frac{1}{N}\langle S_{t'}|S_{t}\rangle=\langle \Psi|\mathbbm{1}_{S}\otimes |t'\rangle\langle t||\Psi\rangle
=\langle \sigma_{t't}^x\rangle+i\langle\sigma_{t't}^y\rangle\,,\end{equation}
where $\sigma_{t't}^x=|t'\rangle\langle t|+|t\rangle\langle t'|$,  $\sigma_{t't}^y=(|t'\rangle\langle t|-|t\rangle\langle t'|)/i$ 
are hermitian Pauli  operators for the pair $t\neq t'$. 

Let us now consider the evolution for a general constant Hamiltonian $H$ of arbitrary spectrum for system $S$, such that 
$U_t=e^{-i Ht}$ $\forall$ $t$. In contrast with (\ref{S}), Eq.\ (\ref{S22}) can in this case be explicitly evaluated. Writing 
\begin{equation}
|S_0\rangle=\sum_k c_k |E_k\rangle,\;\;\;H|E_k\rangle=E_k|E_k\rangle\,,\label{SE}
\end{equation}
with $E_k\neq E_{k'}$ if $k\neq k'$ (in case of degenerate states $|k_l\rangle$,  $c_k|E_k\rangle=\sum_l c_{kl}|k_l\rangle$, 
with $|c_k|^2=\sum_l |c_{kl}|^2$),
then $|S_t\rangle=\sum_k e^{-iE_k t} c_k |E_k\rangle$ and Eq.\ (\ref{S22}) becomes, for equally spaced 
times $t=t_f\frac{j}{N-1}$, $j=0,\ldots, N-1$, 
\begin{eqnarray}E_2(S,T)&=&2(1-\frac{1}{N^2}\sum_{t,t'}|\sum_k |c_k|^2 e^{-iE_k(t-t')}|^2)\\
&=&2\sum_{k\neq k'}|c_k c_{k'}|^2\left[1-\frac{\sin^2\frac{(E_k-E_{k'})t_f N}{2(N-1)}}{N^2\sin^2\frac{(E_k-E_{k'})t_f}{2(N-1)}}\right].\;\;\;\;\;\;\;
\label{E2x}
\end{eqnarray}
 The exact result for a  continuous evolution can also be obtained from (\ref{E2x}), 
 by taking the limit $N\rightarrow\infty$: 
\begin{eqnarray}E_2(S,T)&\underset{N\rightarrow\infty}{\rightarrow}&2\sum_{k\neq k'}|c_k c_{k'}|^2\left[1-\frac{\sin^2\left(\frac{(E_k-E_{k'})t_f}{2}\right)}{(\frac{(E_k-E_{k'})t_f}{2})^2}\right]\;\;\;\;\;\;
\label{E23}
\end{eqnarray}
Eq.\ (\ref{E23})  provides a good approximation to (\ref{E2x}) if $\frac{|E_k-E_{k'}|t_f}{N-1}\ll 1$  $\forall$ $k\neq k'$  
with  finite weight $|c_kc_{k'}|^2> 0$.

Eqs.\ (\ref{E2x})--(\ref{E23}) are essentially measures of the {\it spread} of $|S_0\rangle$ over distinct 
energy eigenstates. For small $t_f$ such that 
$|E_k-E_{k'}|t_f\ll 1$ $\forall\,k,k'$, a second order expansion shows they 
are proportional to the {\it energy fluctuation} in $|S_0\rangle$: 
$|\langle S_{t}|S_{t'}\rangle|^2\approx 1-\langle (\Delta H)^2\rangle(t-t')^2$, 
with $\Delta H=H-\langle H\rangle$ and $\langle O\rangle=\langle S_0|O|S_0\rangle$, implying 
\begin{equation}
E_2(S,T)\approx \frac{N+1}{3(N-1)}\langle (\Delta H)^2\rangle\,t_f^2
\underset{N\rightarrow\infty}{\rightarrow}\frac{1}{3}\langle(\Delta H)^2\rangle\,t_f^2 \,.
\end{equation}
It then becomes proportional to the square of the speed $\sqrt{\langle (\Delta H)^2\rangle}$ 
of the continuous quantum evolution according to the Fubini-Study metric  \cite{AA.90,La.17}. 

It is also apparent from (\ref{E2x}) that $E_2(S,T)$ is upper bounded by the quadratic entropy of the energy distribution $|c_k|^2$: 
\begin{eqnarray}E_2(S,T)\leq 2\sum_{k\neq k'}|c_k c_{k'}|^2=2(1-\sum_k |c_k|^4)\,.
\label{E22}
\end{eqnarray}
The maximum (\ref{E22}) for a fixed distribution $|c_k|^2$ is reached  for an equally spaced  spectrum of the form 
\begin{equation} E_k={\frac{N-1}{t_f}} \frac{2\pi k}{N}+C\,,\label{spec}\end{equation}
with $k$ integer $\in[0,N-1]$,  since in this case the bracket in  (\ref{E2x}) takes its maximum value $1$ $\forall$ $k\neq k'$.   

The spectrum (\ref{spec}) is just Eq.\ (\ref{KST}) for the scaled Hamiltonian 
$H_S=\frac{t_f}{N-1}(H-C)$ (for which $t=0,\ldots,N-1$), so that the 
energy states $|E_k\rangle$ become the Schmidt states $|k\rangle_S$ of (\ref{Scm}) 
and $|c_k|$ the Schmidt coefficients $\lambda_k$. For other spectra, 
the states $|\tilde{k}\rangle_T=
\frac{1}{\sqrt{N}}\sum_t e^{-i E_k t}|t\rangle$ 
in 
\begin{equation}|\Psi\rangle=\frac{1}{\sqrt{N}}\sum_{k,t} c_k e^{-iE_k t}|E_k\rangle |t\rangle=\sum_k c_k|E_k\rangle|\tilde{k}\rangle_T\,,\end{equation}
are not necessarily  all orthogonal, so that $E(S,T)$ will become normally smaller  \cite{BR.16}. 
Nonetheless, for large $N$ and not too small $t_f$, the 
states $|\tilde{k}\rangle_T$ will typically  be almost orthogonal,  so that the deviation from the upper 
bound (\ref{E22}) will not be large, becoming significant only in the presence of quasidegeneracies in the spectrum: 
The bracket in  (\ref{E23}) vanishes just for $E_k\rightarrow E_{k'}$, becoming  close to $1$ for $|E_k-E_{k'}|t_f/2>\pi$, 
while that in (\ref{E2x}), which is a periodic function of $E_k-E_{k'}$ with period $\Delta_N=2\pi\frac{N-1}{t_f}$, 
vanishes for $E_k\rightarrow E_{k'}+m\Delta_N$, $m=0$ or integer,
becoming close to $1$ whenever $|E_k-E_{k'}-m\Delta_N|t_f/2>\pi$.   

On the other hand, Eq.\ (\ref{E23})  also admits a {\it lower bound} for  fixed initial and final states $|S_0\rangle$ 
and $|S_{t_f}\rangle=\sum_k c_k e^{-i E_k t_f}|E_k\rangle$, 
reached when the evolution (over $N$ equally spaced times $t=t_f \frac{j}{N-1}$ under a constant $H$) 
remains in the subspace spanned by $|S_0\rangle$ and $|S_{t_f}\rangle$:
\begin{eqnarray}E_2(S,T)\geq E^{\rm min}_2(S,T) =1-\frac{\sin^2\frac{N\phi}{N-1}}{N^2\sin^2\frac{\phi}{N-1}}\,,
\label{E24}
\end{eqnarray}
where $\phi\in[0,\pi/2]$ is determined by the overlap between the initial and final states:  
\begin{equation}
\cos\phi=|\langle S_0|S_{t_f}\rangle|=|\sum_k |c_k|^2 e^{-iE_k t_f}|\,.\label{ov}
\end{equation}
Writing the final state  as 
\begin{equation}
|S_{t_f}\rangle=e^{-i\gamma}(\cos\phi |S_0\rangle+\sin\phi|S_0^\perp\rangle),
\label{Stf}\end{equation}
where $\langle S_0^\perp|S_0\rangle=0$, $E^{\rm min}_2(S,T)$ is the result of Eq.\ (\ref{E2x}) 
for an evolution under a two level Hamiltonian 
\begin{equation}
H^{\rm min}=\frac{\phi}{t_f}\sigma_y+\frac{\gamma}{t_f}\,,
\;\;\sigma_y=-i(|S_0\rangle\langle S_0^\perp|-|S_0^\perp\rangle\langle S_0|)\,,
\label{Hmin}\end{equation}
such that 
\begin{eqnarray}|S^{\rm min}_t\rangle&\equiv&\exp[-i H^{\rm min} t]|S_0\rangle\nonumber\\&=&
{\textstyle e^{-i\gamma t/t_f}(\cos\frac{\phi t}{t_f}|S_0\rangle+
\sin \frac{\phi t}{t_f}|S_0^\perp\rangle)}\,,\label{Stg}\end{eqnarray}
with  $|S^{\rm min}_{t_f}\rangle=|S_{t_f}\rangle$. 

The demonstration of (\ref{E24}) is given in the appendix, but the result is physically clear: 
The $S-T$ entanglement is a measure of the distinguishability 
between the evolved states, and the minimum value is then obtained for an evolution within the subspace 
containing the initial and final states, 
where all intermediate states will be closer than in a general evolution. 
Such evolution, Eq.\ (\ref{Stg}),  proceeds precisely along the geodesic 
determined by the Fubini-Study metric  \cite{AA.90, La.17}, saturating the Mandelstam-Tamm bound
\cite{Bt.83} 
$\Delta t \Delta E\geq \cos^{-1}(|\langle S_0|S_{t_f}\rangle|)=\phi$ ($\Delta t=t_f$,  
$\Delta E=\sqrt{\langle (\Delta H^{\rm min})^2\rangle}=\phi/t_f$).   

As check, for small $t_f$ such that  $|E_k-\!E_{k'}|t_f\ll 1$ $\forall\, k\neq k'$, 
a fourth order expansion of (\ref{E2x}) and (\ref{E24}) leads to 
\begin{equation}
E_2(S,T)-E^{\rm min}_2(S,T)\approx\kappa[
\langle (\Delta H)^4\rangle-\langle (\Delta H)^2\rangle^2]t_f^4\geq 0\,, \label{46}
\end{equation}
where
$\kappa=\frac{(N+1)(N-2)(N-4/3)}{60(N-1)^3}>0$ $\forall\,N>2$. 
Hence, the difference (\ref{46}) is verified to be non-negative and of fourth order in $t_f$, being proportional 
to the fluctuation of $(\Delta H)^2$. The latter  vanishes just for the geodesic evolution, where 
$\Delta H=\Delta H^{\rm min}=\frac{\phi}{t_f}\sigma_y$ and  hence 
$\langle(\Delta H^{\rm min})^4\rangle=\langle (\Delta H^{\rm min})^2\rangle^2=\phi^4/t_f^4$, 
implying $E_2(S,T)=E^{\rm min}_2(S,T)$. 
Such fluctuation represents a curvature coefficient  which  measures  the deviation from the geodesic \cite{La.17,Br.96}.  

For $\phi\in[0,\pi/2]$, the bound (\ref{E24}) is, of course, an increasing function of $\phi$ for $N\geq 2$, i.e.\  
of the Wootters distance \cite{Wo.81} $s(|S_0\rangle,|S_{t_f}\rangle)=2\arccos(
|\langle S_0|S_{t_f}\rangle|)=2\phi$, and hence a decreasing function of 
the overlap $|\langle S_{t_f}|S_0\rangle|$. It is also a {\it decreasing} function of $N\geq 2$ for  $\phi\in(0,\pi/2]$. 
The minimum value is thus achieved in the continuous limit $N\rightarrow\infty$, 
where $E_2^{\rm min}(S,T)\rightarrow 1-(\sin^2\phi)/\phi^2$.  Then, we may also write, for any $N\geq 2$, 
\begin{eqnarray}
E_2(S,T)\geq 1-\frac{\sin^2\phi}{\phi^2}\,.
\label{E25}
\end{eqnarray}

 \section{Entanglement and history states of evolution operators \label{III}}
 We now examine the application of the previous formalism to the evolution operators themselves. 
 The aim is to link  properties of previous history states with those of the operators that generate it. 
 For this purpose the pure state representation of operators \cite{Ja.72,Ch.75,Gr.05,Du.05,Mi.13}   
 provides a convenient approach, enabling a direct derivation of their entanglement properties \cite{Z.00,N.03,P.07,M.13}.   
\subsection{Entanglement of operators and pure state representation}
We first briefly review  the concept of operator entanglement and its pure state representation. 
Any  operator ${\cal W}$ for a bipartite system A+B can be expanded  as 
\begin{equation}
{\cal W}=\sum_{i,j} M_{ij}C_i\otimes D_j,\label{Ucd}
\end{equation}
where $C_i$ and $D_j$ are orthogonal operators for A and B respectively, 
 satisfying 
\begin{equation}{\rm Tr}\,C_i^\dagger C_j=\delta_{ij}d_A\,,\;\;{\rm Tr}\,D_i^\dagger D_j=\delta_{ij}d_B\,.\label{ort}\end{equation} 
Hence, $M_{ij}=\frac{1}{d_A d_B}{\rm Tr}\,[C_i^\dagger\otimes D_j^\dagger\,{\cal W}]$. We can use, for instance, the 
Weyl operators (\ref{UW}) for the sets $\{C_i\}$, $\{D_i\}$. 

Eqs.\ (\ref{ort}) imply ${\rm Tr}\,[{\cal W}^\dagger {\cal W}]=
d_A d_B{\rm Tr}\,[M^\dagger M]$. If ${\cal W}$ is unitary, then  ${\rm Tr}\,[M^\dagger M]=1$, entailing that the numbers $\{|M_{ij}|^2\}$ 
are in this case standard probabilities. By means of the singular value decomposition, we can write the $d_A^2\times d_B^2$ matrix $M$ as 
$M=UDV^\dagger$,
where $U$ and $V$ are unitary  matrices and $D$ a diagonal matrix with nonnegative entries $\lambda_k^{\cal W}$ 
satisfying $\sum_k (\lambda^{\cal W}_k)^2={\rm Tr}\,M^\dagger M=1$. 
We can then rewrite ${\cal W}$ in the Schmidt form 
\begin{equation}
{\cal W}=\sum_k \lambda^{\cal W}_k A_k \otimes B_k \,,\label{Sf}
\end{equation}
where $A_k \equiv \sum_{i}U_{ik}C_i$ and $B_k \equiv \sum_{j}V^*_{jk}D_j$, are again orthogonal operator bases for $A$ and $B$ satisfying 
${\rm Tr}\,A_k^\dagger A_l=d_A\delta_{kl}$, ${\rm Tr}\,B_k^\dagger B_l=d_B\delta_{kl}$. 
The von Neumann entanglement entropy of ${\cal W}$ can then be defined as  
\begin{equation}
E({\cal W})=-\sum_k (\lambda^{\cal W}_k)^2\log_2 (\lambda^{\cal W}_k)^2\,.\label{EU}
\end{equation}
Similarly,  $E_2({\cal W})=2\sum_k(1- (\lambda_k^{\cal W})^4)$. These entropies vanish when ${\cal W}$ is a product of local 
unitaries, and are maximum when ${\cal W}$ is a uniform sum of $d^2$ products  $A_k\otimes B_k$,  with $d={\rm Min}[d_A,d_B]$.  

The previous analogy between operators and states can be manifestly described 
through the Choi isomorphism \cite{Ja.72,Ch.75,Gr.05,Du.05,Mi.13}. Any operator $O$ in a system with  Hilbert space ${\cal H}$ of 
dimension $d$ can be associated with a pure state $|O\rangle\in{\cal H}\otimes {\cal H}$, given by 
\begin{equation}|O\rangle=(O\otimes \mathbb{1})|\mathbb{1}\rangle=\frac{1}{\sqrt{d}}
\sum_q(O|q\rangle)|q\rangle=\frac{1}{\sqrt{d}}\sum_{q,q'}\langle q'|O|q\rangle|q'\rangle|q\rangle\,,
\end{equation}
where $|\mathbb{1}\rangle=\frac{1}{\sqrt{d}}\sum_q |q\rangle |q\rangle$ is a maximally entangled state in ${\cal H}\otimes {\cal H}$ and $\{|q\rangle\}$ 
an orthonormal set. In this way, 
\begin{equation}\langle O|O'\rangle=\frac{1}{d}\rm Tr\,[O^\dagger O']\,.\end{equation} 
Therefore,  orthogonal operators satisfying ${\rm Tr}\,[O_i^\dagger O_j]=d\delta_{ij}$ correspond to orthonormal states $\langle O_i|O_j\rangle=\delta_{ij}$. 
And unitary operators $U$ to normalized states $|U\rangle$. 

The operator (\ref{Ucd})  can then be associated with the pure state (note   that $|\mathbbm{1}_{AB}\rangle=|\mathbbm{1}_A\rangle|\mathbbm{1}_B\rangle$)
\begin{equation}
|{\cal W}\rangle=({\cal W}\otimes\mathbbm{1}_{A'B'}) |\mathbbm{1}_A\rangle|\mathbbm{1}_B\rangle
=\sum_{ij} M_{ij}|C_i\rangle |D_j\rangle\,,\label{StU}
\end{equation}
where $|C_i\rangle=(C_i\otimes \mathbbm{1}_{A'})|\mathbbm{1}_A\rangle$, 
 $|D_j\rangle=(D_j\otimes \mathbbm{1}_{B'})|\mathbbm{1}_B\rangle$ form  orthogonal sets:  
 $\langle C_k|C_i\rangle=\delta_{ki}$, 
 $\langle D_k|D_j\rangle=\delta_{kj}$. Thus, $M_{ij}=\langle C_i,D_j|{\cal W}\rangle$, 
 with $\langle {\cal W}|{\cal W}\rangle={\rm Tr}\,[M^\dagger M]$. 

The state representation of the Schmidt form (\ref{Sf}) acquires then the standard appearance 
\begin{equation}|{\cal W}\rangle=\sum_k \lambda^{\cal W}_k |A_k\rangle|B_k\rangle\,,\end{equation}
with $\langle A_k|A_l\rangle=\delta_{kl}=\langle B_k|B_l\rangle$, and the entanglement entropy (\ref{EU}) of a unitary 
${\cal W}$ can be also expressed as 
\begin{equation}
E({\cal W})=S(\rho_A^{\cal W})=S(\rho_B^{\cal W})\,,\;\;
\rho_{A(B)}^{\cal W}={\rm Tr}_{B(A)}\,|{\cal W}\rangle\langle {\cal W}|\,,\label{EU2}
\end{equation}
with $S(\rho)=-{\rm Tr}\rho\log_2\rho$.  Similarly, 
$E_2({\cal W})=S_2(\rho_A^{\cal W})=S_2(\rho_B^{\cal W})$, with 
$S_2(\rho)=2(1-{\rm Tr}\,\rho^2)$. 

\subsection{Generating operators and operator history  states}
The history state (\ref{St1}) can be generated from an initial product state $|S_0\rangle|0\rangle$ as 
\begin{equation}
|\Psi\rangle={\cal W}(I\otimes H^{\otimes n})|S_0\rangle|0\rangle\,, \label{CW}
\end{equation}
where $H^{\otimes n}$ denotes the Hadamard operator acting on the clock 
($H^{\otimes n}|0\rangle=\frac{1}{\sqrt{N}}\sum_{t=0}^{N-1}|t\rangle$, with $N=2^n$) and  
\begin{equation}
{\cal W}=\sum_t U_t\otimes |t\rangle\langle t|,\label{U1}
\end{equation}
the control-$U_t$ operator. By expanding $U_t$ in an orthogonal basis of operators $C_i$, we have  
\begin{equation}{\cal W}=\sum_{t,i} {M}_{ti} C_i \otimes |t\rangle\langle t|,\;\;
{M}_{ti}=\frac{1}{d_S}{\rm Tr}\,C_i^\dagger U_t\,,\label{U12}
\end{equation}
where the coefficients ${M}_{tj}$ satisfy $\sum_j |{M}_{tj}|^2 =\frac{1}{d_S}{\rm Tr}\,U_t^\dagger U_t=1$, 
and are hence standard probabilities at fixed $t$. 
Since the projectors $|t\rangle\langle t|$ are also orthogonal and have unit trace, 
the Schmidt coefficient $\lambda_k^{\cal W}$ of (\ref{Sf}) are  here just the singular values of the matrix ${M}/\sqrt{N}$. 
The ensuing entanglement entropy (\ref{EU}) is the same as that of  ${\cal W}(I\otimes H^{\otimes n})$, 
as they  differ just by a local unitary.  

\begin{figure}[h!]
 \centering
\includegraphics{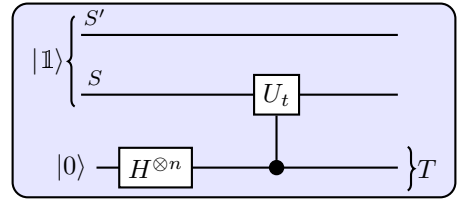}
 \caption{(Color online) Schematic circuit representing the generation of the operator history state (\ref{UH}). } \label{f3}
\end{figure}

The pure state (\ref{StU}) associated with the operator (\ref{U1}) 
is itself an {\it operator history state}: 
\begin{equation}
|{\cal W}\rangle=\frac{1}{\sqrt{N}}\sum_t 
|U_t\rangle|T_t\rangle\label{UH}\,,
\end{equation}
where $|U_t\rangle=(U_t\otimes\mathbbm{1}_{S'})|\mathbbm{1}_S\rangle=\frac{1}{\sqrt{d_S}}\sum_q U_t|q\rangle|q\rangle$ 
and $|T_t\rangle=(T_t\otimes\mathbbm{1}_{T'})|
\mathbbm{1}_T\rangle=|tt\rangle$, with $T_t=\sqrt{N}|t\rangle\langle t|$ 
 and $\langle T_t|T_{t'}\rangle=\delta_{tt'}$. 
 Writing $|tt\rangle$ simply as $|t\rangle$, Eq.\ (\ref{UH}) is the standard history state (\ref{St1}) for a 
 maximally entangled initial  state $|\mathbbm{1}_{S}\rangle=\frac{1}{\sqrt{d_s}}\sum_q|q\rangle|q\rangle$ 
 of a bipartite system under a local evolution 
 $U_t\otimes \mathbbm{1}_{S'}$, so that it can be generated with the circuit depicted in Fig.\ \ref{f3}. 

The entanglement of the history state (\ref{UH}) is the operator entanglement  (\ref{EU}) of ${\cal W}$, which is then a measure  
of the distinguishability of the operator states $|U_t\rangle$. Its quadratic entanglement  
can be directly evaluated with Eq.\ (\ref{S22}), where now $\langle U_t|U_{t'}\rangle=\frac{1}{d_S}{\rm Tr}\,[U_t^\dagger U_{t'}]$:
\begin{equation}
    E_2({\cal W})=2(1-{\textstyle\frac{1}{N^2}}\sum_{t,t'}|\langle U_t|U_{t'}\rangle|^2)\,.\label{E2W}
\end{equation} 

It is now immediate to see that if $N=d_S^2$ and the operators $\{U_t\}$ form a {\it complete orthogonal set} 
(Eq.\ (\ref{Uto})), the operator history state (\ref{UH}) is {\it maximally entangled}:  
\begin{equation}
E({\cal W})=\log_2 d_S^2=2\log_2 d_S\,,
\end{equation}
while $E_2({\cal W})=2(1-\frac{1}{d_S^2})$, 
since all states $|U_t\rangle$ become orthogonal: $\langle U_t|U_{t'}\rangle=\delta_{tt'}$. 
The form (\ref{UH}) is then already the Schmidt decomposition of $|{\cal W}\rangle$.  Since in this case 
the original history state (\ref{St1}) has maximum entanglement $E(S,T)=\log_2 d_S$ for any initial state 
$|S_0\rangle$, this  result indicates a 
close relation between the  entangling power of ${\cal W}$ and its operator entanglement, which will be discussed below. 
It is also apparent that if the $d_S^2$ operators $U_t$ are not all orthogonal, then $E({\cal U})<2\log_2 d_S$.  

For a smaller number $N< d_S^2$ of times, $E({\cal W})$ will be maximum if all $N$  states $|U_t\rangle$ are orthogonal. 
In the case of a {\it constant} Hamiltonian  with energies $E_k$, such that  $U_t=e^{-iHt}$ $\forall$ $t$, then 
\begin{equation}\langle U_t|U_{t'}\rangle=\frac{1}{d_S}\sum_k e^{-iE_k(t-t')}\,.\end{equation}
For $N=d_S$, an equally spaced  spectrum  $E_k=2\pi k/N+C$, $k=0,\ldots,N-1$, (i.e., Eq.\  (\ref{spec}) 
if $t\rightarrow\frac{t_f}{N-1}j$) ensures that all $|U_t\rangle$ are strictly orthogonal: 
$\langle U_t|U_{t'}\rangle=\delta_{tt'}$ $\forall$ $t,t'$ (the ensuing operators $U_t$  are in fact the first $d_S$ operators 
of the Weyl set (\ref{UW})). Hence, $E({\cal W})$ will reach for this spectrum the maximum value 
\begin{equation}E({\cal W}) =\log_2 d_S\,,\end{equation}
compatible with a fixed $H$ and $N=d_S$ times.  The same holds for $E_2({\cal W})$. This result correlates 
with the extremal properties of this spectrum discussed in  \ref{IIc}. 

On the other hand, since $U_{t,t-1}=U_tU_{t-1}^\dagger$, the operator ${\cal U}$ of Eq.\ (\ref{Upsi}) 
is related with ${\cal W}$ by
\begin{equation} 
{\cal U}={\cal W}(I\otimes \exp[-iP_T]){\cal W}^\dagger\,,
\end{equation}
where $\exp[-iP_T]=\sum_t |t\rangle\langle t-1|$.  The associated pure state is also a history state, 
\begin{equation}
    |{\cal U}\rangle=\frac{1}{\sqrt{N}}\sum_t |U_{t,t-1}\rangle|T_{t,t-1}\rangle\,,\label{Utm1}
\end{equation}
where $|T_{t,t-1}\rangle=\sqrt{N}(|t\rangle \langle t-1|\otimes \mathbb{1}_{T'})|\mathbb{1}_T\rangle=|t,t-1\rangle$ are again orthogonal states. 
Its entanglement is then a measure of the distinguishability of the step evolution operator states $|U_{t,t-1}\rangle$, 
and depends on the {\it order} of the operators $U_t$, in contrast with $E({\cal W})$. It vanishes in the constant case (\ref{UHS})--(\ref{J}). 

\subsection{Operator entanglement and entangling power \label{IIIc}}
We have seen that there is a relation between the entanglement  of the operator ${\cal W}$ and that of the history states it generates, 
  $|\Psi\rangle=\frac{1}{\sqrt{N}}\sum_{t}U_t|S_0\rangle |t\rangle$.
 We will here  prove that the quadratic operator entanglement entropy $E_2(U,T)\equiv E_2({\cal W})$, Eq.\ (\ref{E2W}), is 
  proportional to the {\it entangling power} of ${\cal W}$, defined  as the average quadratic entanglement it generates when applied (as in Eq.\ (\ref{CW}))  to  initial product states $|S_0\rangle|0\rangle$:  
\begin{equation}
 \langle E_2 (S,T)\rangle=\frac{d_S}{d_S+1}E_2({\cal W})\,, \label{rel}
\end{equation}
where  
 \begin{equation}
 \langle E_2 (S,T)\rangle=
 \int_{\cal H} 2(1- {\rm Tr}\,\rho_S^2) d S_0\,, \label{ME2}
\end{equation}
is the average over all initial states $|S_0\rangle$ of the quadratic entanglement entropy  $E_2(S,T)$ of the history state:  
The integral runs over the whole set of initial states $|S_0\rangle$ with the Haar measure $dS_0$ (the only normalized unitarily 
invariant measure over the Hilbert space) and $\rho_S$ is the reduced state of $S$ in $|\Psi\rangle$. \\ 
{\it Proof.} Since $\rho_S=\frac{1}{N}\sum_t U_t|S_0\rangle \langle S_0|U_t^\dagger$, we obtain 
\begin{equation}
\langle {\rm Tr}\,\rho_S^2\rangle= 
 \frac{1}{N^2}\sum_{t,t'} \int_{\cal H}\langle S_0| U_t^{\dagger}U_{t'}|S_0\rangle\langle S_0| U_{t'}^{\dagger}U_t|S_0\rangle d S_0\,.
  \label{ME21}
\end{equation}
Here we can define $O=U_t^{\dagger}U_{t'}$ and $P=U_{t'}^{\dagger}U_t=O^\dagger$ to use the relation \cite{RBKSC.04} 
\begin{equation}
 \int_{{\cal H}} \langle S_0|O|S_0\rangle\langle S_0|P|S_0\rangle d S_0=\frac{{\rm Tr}[O] {\rm Tr}[P]+{\rm Tr}[OP]}{d_S(d_S+1)} \,.
 \label{MEtr}
\end{equation}
Since in this case  $OP=\mathbb{1}_S$, we obtain
\begin{equation}
\langle {\rm Tr}\,\rho_S^2\rangle= \frac{\frac{1}{N^2}\sum_{t,t'} |{\rm Tr}\,[U_t^{\dagger}U_{t'}]|^2 + d_S}{d_S(d_S+1)}. \label{ME22}
\end{equation}
On the other hand, $E_2 ({\cal W})=2(1-{\rm Tr}\,\rho_U^2)$, 
with $\rho_U^2=\frac{1}{N^2}\sum_{t,t'} |U_t\rangle\langle U_t^{\dagger}|U_{t'}\rangle\langle U_{t'}^{\dagger}|$. Thus, 
\begin{equation}
{\rm Tr}\,\rho_U^2
=\frac{1}{N^2}\sum_{t,t'} |\langle U_t^{\dagger}|U_{t'}\rangle|^2  = 
\frac{1}{(d_S N)^2}\sum_{t,t'} |{\rm Tr} [U_t^{\dagger}U_{t'}]|^2\,. \label{TrUt}
\end{equation}
Replacing (\ref{TrUt}) in (\ref{ME22}) leads to  
$\langle {\rm Tr}\,\rho_S^2\rangle=\frac{d_S{\rm Tr}\,(\rho_U^2)+1}{d_S+1}$ and hence to Eq.\ (\ref{rel}). \qed 

Therefore, the average over all initial system states of the quadratic $S-T$ entanglement is just that of the 
generating unitary operator times $\frac{d_S}{d_S+1}$. 
It is first verified that if the operators $U_t$ form a complete orthogonal set, $E_2({\cal W})=2(1-d_S^{-2})$ 
is maximum and Eq.\ (\ref{rel}) yields $\langle E_2(S,T)\rangle=2(1-d_S^{-1})$,  	
 the maximum attainable value in a $d_S$ dimensional space,  entailing it  is always maximum, irrespective of the 
 initial state  (sec.\ \ref{IIbb}).  

In general, for a reduced set of $d$ orthogonal unitaries 
$U_t$, with $N=d\leq d_S^2$, $E_2({\cal W})=2(1-d^{-1})$ and  hence
\begin{equation}
    \langle E_2(S,T)\rangle=2\frac{d_S(d-1)}{d(d_S+1)}\,.\label{rel2}
\end{equation}
 In order to visualize this relation we define the effective  average number of orthogonal states the system 
 visits as
 \begin{equation}
     \overline{d}_{S,T}=\frac{1}{1-\frac{1}{2}\langle E_2(S,T)\rangle}=\frac{d(d_S+1)}{d_S+d}\label{rel3}\,,
\end{equation}
such that  $\langle E_2(S,T)\rangle=2(1-\frac{1}
{\overline{d}_{S,T}})$. If $d=d_S^2$,    $\overline{d}_{S,T}=d_S$ becomes maximum, while if $d=d_S$, 
which is, for instance, the case of a constant 
Hamiltonian with spectrum $2\pi k/N$ ($d_S$ orthogonal operators $U_t=\exp[-iHt]$), Eq.\ (\ref{rel3}) 
leads to $\overline{d}_{S,T}=(d_S+1)/2$, i.e., 
just {\it half} the maximum value for large $d_S$. For any other spectrum and $N=d_S$,  $\overline{d}_{S,T}\leq (d_S+1)/2$, i.e., 
\begin{equation}
     \langle E_2(S,T)\rangle\leq 2\frac{d_S-1}{d_S+1}\label{rel4}\;\;\;\;\;(U_t=e^{-iHt},\;\;N=d_S)\,.
\end{equation}
Noticeably, it is sufficient to have  $d\propto d_S$ ($\ll d_S^2$ for large $d_S$) to reach a high $\overline{d}_{S,T}$, i.e., 
$\overline{d}_{S,T}=\frac{m}{m+1}(d_S+1)$ if $d=m d_S$ (and $m\leq d_S$), as seen from (\ref{rel3}). 

\subsection{Measuring operator overlaps} 
\begin{figure}[h!]
 \centering
  \includegraphics{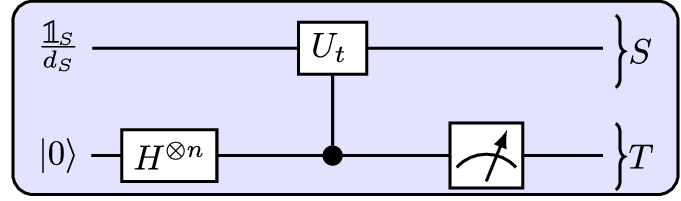}
 \caption{(Color online) Schematic circuit representing the measurement of the operator overlaps (\ref{Uov}).} \label{f4}
\end{figure}

The overlaps $\langle U_t|U_t'\rangle$, 
which are the operator fidelities defined in \cite{Wa.09} and are involved in the quadratic entanglement 
(\ref{E2W}) of the generating operator ${\cal W}$,  can be experimentally obtained  by measuring 
$|T_t\rangle\langle T_{t'}|$ in the time part $T$ (Fig.\ \ref{f4}). 
Remarkably, it is sufficient to start with the system in a  {\it maximally mixed} state: If we trace out system $S'$ 
in the operator history state (\ref{UH}), we obtain  
\begin{equation}\rho_{ST}=\frac{1}{N d_S}\sum_{t,t'}U_t U_{t'}^{\dagger}\otimes|t\rangle\langle t'|\label{rst}\,,\end{equation}
where we have written $|T_t\rangle$ as $|t\rangle$. Hence, tracing over $S$, 
\begin{equation}\rho_{T}=
\frac{1}{N}\sum_{t,t'}\langle U_{t'}| U_{t}\rangle|t\rangle\langle t'|\,.\end{equation} 
Thus, setting $U_{t,t'}=U_t U_{t'}^\dagger$, 
\begin{equation}
\langle|t'\rangle \langle t|\rangle=\frac{1}{N}\langle U_{t'}|U_{t}\rangle =
\frac{1}{N d_S}{\rm Tr\,}[U_{t'}^\dagger U_{t}]=\frac{1}{N d_S}{\rm Tr\,}[U_{t,t'}]\,.\label{Uttp}
\end{equation}
Using again $\sigma^x_{t't}=
|t'\rangle\langle t|+|t\rangle\langle t'|$, $\sigma^y_{t't}=-i(|t'\rangle\langle t|-|t\rangle\langle t'|)$,  
the trace of the evolution operator between any two times can then be obtained  by measuring the averages of 
 $\sigma_{tt'}^x$ and $\sigma^y_{t't}$, which provide the real and imaginary parts: 
 \begin{equation}
\langle \sigma^x_{t't}\rangle=
\frac{2}{N}{\rm Re}[\langle U_t|U_{t'}\rangle],\;\;\langle \sigma^y_{t't}\rangle=
\frac{2}{N}{\rm Im}[\langle U_t|U_{t'}\rangle]\,.\label{Uov}
    \end{equation}

Of course, the state  (\ref{rst}))  can be generated  just by preparing system $S$  in the maximally mixed state, 
as the purifying system $S'$ of the original operator state is traced out. Note also that $U_0=\mathbbm{1}$, 
so that the averages $\langle\sigma^\mu_{0t}\rangle$ determine ${\rm Tr}\,U_t$.  

In the special case  $N=2$,  
$T$ is a single qubit and we recover the standard DQC1 scheme  for measuring the trace of an operator \cite{KL.98}. 
 The ensuing operator history state is 
 $|{\cal W}\rangle=
 \frac{1}{\sqrt{2}}(|U_0\rangle|T_0\rangle+|U_1\rangle|T_1\rangle)$,  and its quadratic entanglement is
 \begin{equation}
 E_{2}({\cal W})=1-|\langle U_0|U_1\rangle|^2=1-|{\rm Tr}\,U|^2/d_S^2\,.
\end{equation}
 Its square root is just the entangling power of the DQC1 circuit  defined in \cite{Yi.13}. 
 
\section{Conclusions\label{IV}}

Quantum mechanics has mostly considered time as an external classical parameter.
In this work we have determined some fundamental properties concerning the generation and entanglement of discrete history states within 
a parallel-in-time discrete model of quantum evolution, based on a finite dimensional quantum clock \cite{BR.16}. 
It was first shown that a general unitary evolution for the system states follows from a static eigenvalue equation,  
which can be recast as a generalized discrete version of a Wheeler-DeWitt equation. 
The ensuing system-time entanglement is a measure of the actual number of distinguishable states visited by the system 
at distinguishable times and satisfies an entropic energy-time uncertainty relation. Its dependence  on  the  initial system 
state becomes attenuated for non-constant non-commuting Hamiltonians, and in particular  we have  presented a simple two-clock scheme which 
generates a {\it maximally entangled} history state irrespective of the seed state. Thus, history states essentially independent 
of initial conditions can be generated. On the other hand, for any fixed seed system state there is always a 
special clock basis selection for which the evolution corresponds to a constant Hamiltonian.

We have also shown that the quadratic entropy provides a convenient measure of the system-time entanglement entropy. It can be 
evaluated analytically and satisfies strict and physical upper and lower bounds,  the former 
connected with the energy spread of the initial state and the latter determined by the 
evolution along the geodesic path between the initial and final states. Hence, such path, which provides the minimum evolution 
time \cite{Bt.83},  minimizes as well the  quadratic  $S-T$ entanglement entropy. 

Finally, by means of the channel-state duality we have shown that  the unitary operator generating the history state corresponds to an operator history state, with  its quadratic entanglement entropy  representing its entangling power.  
We have also provided a simple scheme which allows to efficiently obtain the overlaps between system states and the traces of the
evolution operator between any two-times through measurements on the clock.  

The present formalism is interesting as a fundamental aspect  of quantum theory, where there are some possible 
scenarios to explore further in connection with quantum gravity, such as interaction between relational clocks 
\cite{Bo.11,Ho.12} and emergence of causality \cite{Br.11,CR.17}. 
The incorporation of time in a discrete quantum clock system also enables the development of new models 
of parallel-in-time simulation, taking advantage of the quantum features of superposition and entanglement. 
This description of time could be also suitable for
 application in Floquet systems, and in particular Floquet time crystals \cite{El.16}. 

\section{Appendix}
{\it Proof of the lower bound of Eq.\ (\ref{E24})}. 
 We first assume a sufficiently short final time $t_f$ such that $|\frac{(E_k-E_{k'})t_f}{2}|\leq \pi$ $\forall$ $k\neq k'$. 
 Note that the overlap $|\langle S_0|S_{t_f}\rangle|$, Eq.\ (\ref{ov}), is unaffected by any translation 
 $E_k\rightarrow E_k+2j\pi/t_f$ $\forall j\in\mathbb{Z}$, for a given $k$.
 The angle $\phi\in[0,\pi/2]$ determined by this  overlap can also be rewritten as 
 \begin{eqnarray}
 \phi&=&\arcsin\sqrt{1-|\langle S_0|S_{t_f}\rangle|^2}\\
 &=&\arcsin\sqrt{\textstyle 2\sum_{k\neq k'}|c_k c_{k'}|^2\sin^2\frac{(E_k-E_{k'})t_f}{2}}\,.\label{ov2}
 \end{eqnarray}
 It is now expected that the overlap between any pair of intermediate states will be smaller than those between states 
 $|S^{\rm min}_t\rangle=e^{-iH_{\rm min}t}|S_0\rangle$ 
 along the geodesic, such that (Eq.\ (\ref{Stg}))  
 $|\langle S_t|S_{t'}\rangle|\leq |\langle S^{\rm min}_t|S^{\rm min}_{t'}\rangle|=|\cos[\phi\frac{t-t'}{t_f}]|$. 
 This inequality is verified since the function 
 \begin{equation}
 F(s)=\arcsin\sqrt{\textstyle2\sum_{k\neq k'}|c_k c_{k'}|^2\sin^2\frac{(E_k-E_{k'})t_fs}{2}}-\phi s
\end{equation} 
where $s=|\frac{t-t'}{t_f}|\leq 1$,  is a concave function of $s$ for $s\in [0,1]$ and  satisfies $F(0)=F(1)=0$, 
so that  $F(s)\geq  0$ $\forall$ $s\in [0,1]$. 
Hence, for short times $t_f$ such that  all relative phases have yet not completed one period  
($|E_k-E_{k'}|t_f<2\pi$ $\forall$ $k,k'$), all intermediate overlaps of the actual evolution are smaller than 
those along the geodesic, and hence the actual $E_2(S,T)$ entropy is larger than that along the geodesic path. 

For larger times $t_f$, the inequality (\ref{E24}) also holds but for a different reason: If $|\frac{(E_k-E_{k'})t_f}{2}|>\pi$ 
for some pairs $k,k'$, $F(s)$ may not be concave and can also be negative for some values of $s$. However, the relevant term of 
the exact expression for $E_2(S,T)$ satisfies 
\begin{equation} 
\frac{\sin^2\frac{\gamma N}{N-1}}{N^2\sin^2\frac{\gamma}{N-1}}\leq  
\frac{\sin^2\frac{(\gamma-j\pi) N}{N-1}}{N^2\sin^2\frac{(\gamma-j\pi)}{N-1}}
\end{equation}
where $\gamma=\frac{(E_k-E_{k'})t_f}{2}$ and $j$ is such that $|\gamma-j\pi|\in [0,\pi/2]$. This translation 
of the energy difference does not affect the 
overlap (Eq.\ (\ref{ov2})), but shows that the actual entropy  $E_2(S,T)$ for large times will not become lower 
than the bound previously obtained. In this case 
some relative phases may have completed one or more periods, but the final effect will be to decrease the average 
overlap and hence to  increase  $E_2(S,T)$. 

\acknowledgments
The authors acknowledge support from CIC (RR) and CONICET (AB) of Argentina, and CONICET Grant PIP 112201501-00732.

\end{document}